\documentclass[conference]{IEEEtran}

\usepackage[table]{xcolor}
\usepackage{booktabs}
\usepackage{graphicx}
\usepackage{balance}
\usepackage{comment}
\usepackage{amsmath}
\usepackage{amssymb}
\usepackage[hidelinks]{hyperref}

\usepackage{footnote}
\makesavenoteenv{tabular}
\makesavenoteenv{table}

\usepackage{siunitx}
\DeclareSIUnit\sample{S}
\DeclareSIUnit[group-minimum-digits=5]\usd{USD}

\usepackage[style=numeric,citestyle=numeric-comp,backend=biber,sorting=none,maxbibnames=99]{biblatex}
\newcommand{\sysname}[0]{\textsc{SatIQ}}

\begin{document}

\title{Sticky Fingers: Resilience of Satellite Fingerprinting against Jamming Attacks}

\makeatletter
\newcommand{\linebreakand}{%
  \end{@IEEEauthorhalign}
  \hfill\mbox{}\par
  \mbox{}\hfill\begin{@IEEEauthorhalign}
}
\makeatother

\author{\IEEEauthorblockN{Joshua Smailes}
\IEEEauthorblockA{University of Oxford\\
joshua.smailes@cs.ox.ac.uk}
\and
\IEEEauthorblockN{Edd Salkield}
\IEEEauthorblockA{University of Oxford\\
edd.salkield@cs.ox.ac.uk}
\and
\IEEEauthorblockN{Sebastian K{\"o}hler}
\IEEEauthorblockA{University of Oxford\\
sebastian.kohler@cs.ox.ac.uk}
\linebreakand
\IEEEauthorblockN{Simon Birnbach}
\IEEEauthorblockA{University of Oxford\\
simon.birnbach@cs.ox.ac.uk}
\and
\IEEEauthorblockN{Martin Strohmeier}
\IEEEauthorblockA{armasuisse Science + Technology\\
martin.strohmeier@armasuisse.ch}
\and
\IEEEauthorblockN{Ivan Martinovic}
\IEEEauthorblockA{University of Oxford\\
ivan.martinovic@cs.ox.ac.uk}}

\IEEEoverridecommandlockouts
\makeatletter\def\@IEEEpubidpullup{6.5\baselineskip}\makeatother
\IEEEpubid{\parbox{\columnwidth}{
    {\fontsize{7.5}{7.5}\selectfont Workshop on Security of Space and Satellite Systems (SpaceSec) 2024 \\
    1 March 2024, San Diego, CA, USA \\
    ISBN 979-8-9894372-1-4 \\
    https://dx.doi.org/10.14722/spacesec.2024.23054 \\
    www.ndss-symposium.org}
}
\hspace{\columnsep}\makebox[\columnwidth]{}}

\maketitle

\begin{abstract}

In the wake of increasing numbers of attacks on radio communication systems, a range of techniques are being deployed to increase the security of these systems.
One such technique is radio fingerprinting, in which the transmitter can be identified and authenticated by observing small hardware differences expressed in the signal.
Fingerprinting has been explored in particular in the defense of satellite systems, many of which are insecure and cannot be retrofitted with cryptographic security.

In this paper, we evaluate the effectiveness of radio fingerprinting techniques under interference and jamming attacks, usually intended to deny service.
By taking a pre-trained fingerprinting model and gathering a new dataset in which different levels of Gaussian noise and tone jamming have been added to the legitimate signal, we assess the attacker power required in order to disrupt the transmitter fingerprint such that it can no longer be recognized.
We compare this to Gaussian jamming on the data portion of the signal, obtaining the remarkable result that transmitter fingerprints are still recognizable even in the presence of moderate levels of noise.
Through deeper analysis of the results, we conclude that it takes a similar amount of jamming power in order to disrupt the fingerprint as it does to jam the message contents itself, so it is safe to include a fingerprinting system to authenticate satellite communication without opening up the system to easier denial-of-service attacks.

\end{abstract}

\section{Motivation}\label{sec:motivation}

Due to the growing availability of cheap Software-Defined Radio (SDR) hardware combined with increasing global reliance upon satellite systems and their data, attacks on space systems are a real threat to critical infrastructure.
Furthermore, the great cost of launching and operating satellites means there are a large number of legacy satellite systems lacking proper cryptographic security.
A number of techniques have been proposed to secure satellite communications in the absence of cryptography, relying upon analysis of the signal and other factors to authenticate messages.
Fingerprinting is one of these techniques, and involves looking at the signal impairments caused by small differences in the transmitter hardware to authenticate the transmitter and distinguish it from attacker-controlled devices.
This has been well-explored in terrestrial radio systems, and has seen use more recently in satellite systems, which have the additional difficulty of high levels of atmospheric noise.

However, an adversary's goal is not always to spoof communication; sometimes simple denial of service is sufficient.
Traditionally this has been achieved through jamming techniques, involving the use of targeted noise or other signals to stop the legitimate signal from being properly decoded. This has been observed widely in the real world, particularly in the recent jamming attacks on Starlink in the Ukraine war \cite{ren2023anti,starlinkjamming2}.

If fingerprinting techniques are used to secure a ground system, incoming messages may also be rejected if the transmitter fingerprint does not match the expected value.
Therefore, an attacker may achieve easier denial of service by simply disrupting the fingerprint.
With satellite systems, this is a particular concern -- the high levels of atmospheric noise have already distorted the signal such that any identifiable transmitter characteristics are difficult to extract, which may make the attacker's goal even easier to accomplish.

\subsection{Contributions}\label{sec:contributions}

\begin{figure*}
    \centering
    \includegraphics[width=.8\linewidth]{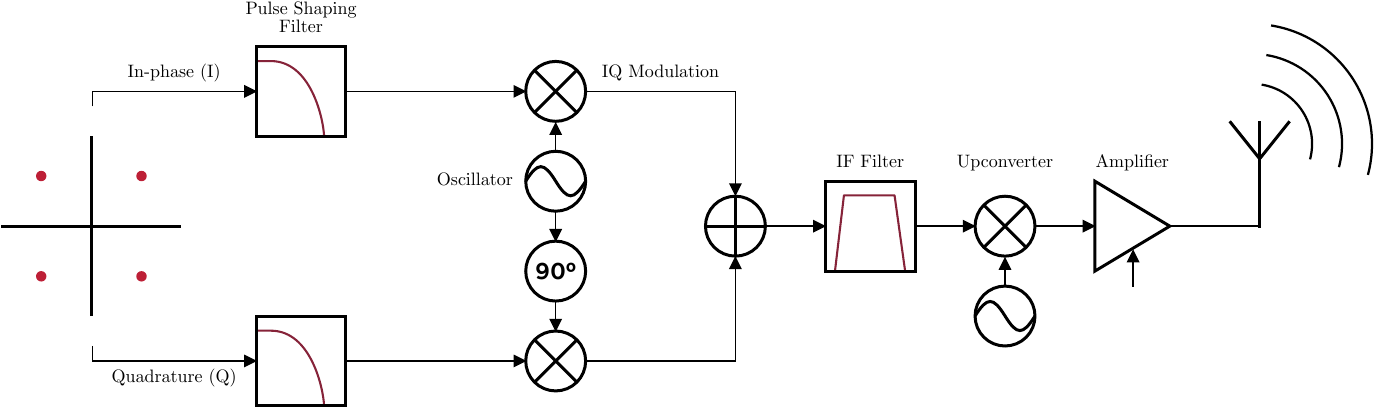}
    \caption{An illustration of the hardware components involved in QPSK signal modulation. Each hardware component can introduce its own impairments upon the signal, which can be used to identify the transmitter through fingerprinting.}
    \label{fig:signal-modulation}
\end{figure*}

In this paper we show that physical-layer satellite fingerprinting techniques are resilient against jamming attacks, so their use does not increase the threat of denial-of-service attacks.
We build upon our previous work in satellite fingerprinting, using the \sysname{} system as a case study~\cite{smailes2023watch}.
We focus on the Iridium satellite constellation used for communication, first performing an analysis of the level of Gaussian noise necessary to disrupt messages and the hardware required to achieve this.
This is followed by data collection, in which we collect \num{540066} messages with varying levels of noise added to the incoming signal, and the creation of further datasets by adding synthetic jamming to clean signals in software.\footnote{The dataset has been made openly available at \url{https://zenodo.org/records/10678124}, and the code at \url{https://github.com/ssloxford/SatIQ-noise}.}
We next perform an evaluation of this data, showing that in order to disrupt the transmitter fingerprint through jamming, a similar or greater transmit power is required than to disrupt the message contents via traditional jamming techniques, with a difference of between \num{0.3} and \qty{2}{\decibel}.
Finally, we conclude with a discussion of these results and their potential root causes.

\section{Background}\label{sec:background}

In this section we provide an overview of radio fingerprinting techniques, the jamming approaches that we evaluate, and the message structure and error correcting code attributes of the target protocol.

\subsection{Fingerprinting}\label{sec:background-fingerprinting}

Fingerprinting techniques are used to identify radio transmitters by looking solely at the received signal.
This is possible due to small variations in the analog components of the transmitter hardware (illustrated in Figure~\ref{fig:signal-modulation}) introducing different impairments onto the signal, even between identical copies of the same hardware~\cite{soltaniehReview2020}.
These impairments can be used to identify or authenticate transmitters, protecting against spoofing and replay attacks by attacker-controlled devices.
This is the main use case for fingerprinting in satellites, many of which are not secured with cryptography for various reasons. Examples include NASA's MODIS data which is open by design~\cite{nasaFirms}, GEO-KOMPSAT-2A with leaked keys~\cite{xrit-rx}, and COMS-1 which has broken cryptography~\cite{lrit-key-dec}.

There are two main approaches to fingerprinting.
Transient fingerprinting techniques look at the signal transient: the portion of the signal where the transmitter first powers up.
This requires precise timing, but has been demonstrated to contain sufficient information for authenticating transmitters in security contexts~\cite{rasmussenImplications2007}.
On the other hand, steady-state fingerprinting looks at the signal as a whole, pulling identifying information out of the modulated portion of the signal~\cite{foruhandehSpotr2020}.

In both steady-state and transient fingerprinting, the fingerprint is also affected by properties of the wireless channel: background noise (including interference from legitimate sources), free space path loss, and multipath distortion all add additional impairments onto the signal, making fingerprint extraction more difficult.
This is particularly prominent in satellite systems, in which signals travel through hundreds of kilometers of atmosphere before reaching the receiver.
A number of approaches have been explored to mitigate this issue, including averaging multiple messages to eliminate noise~\cite{wangRadio2022,oligeriPASTAI2020}, adding noise to clean signals during training~\cite{tekbasImprovement2004}, and looking at high sample rate signals to capture more detail and a wider range of features~\cite{smailes2023watch}.

\subsection{Jamming}\label{sec:background-jamming}

In a jamming scenario, the attacker emits an interfering signal in the same frequency band as the victim signal, within the vicinity of a victim receiver.
Denial of service is achieved once the received power of the attacker's signal causes sufficient bit errors, making a certain proportion of packets undecodable.
The received power of the interfering signal emitted by the attacker can be measured relative to the victim signal received power, giving the attacker-to-victim power ratio.

Against a fingerprinting system, the attacker's objective is to cause victim messages to be rejected as illegitimate, by inducing fingerprinter error on a certain proportion of packets.
The robustness of the fingerprinter against a jammer is therefore seen by comparing the decoder error rate to the fingerprinter error rate, as the attacker-to-victim power ratio increases.

In this work, we consider two main types of jamming which are widely used in recent analyses of space systems: noise jamming and tone jamming~\cite{rawlins2022death,lefcourt2022space}.
It is known that these forms of jamming cause denial of service at different attacker-to-victim power ratios; we evaluate both as we are interested to observe the different effects that they cause in the fingerprinting system.

In particular, we focus on the Iridium satellite constellation, used primarily for communication and composed of 66 satellites in Low Earth Orbit (LEO).
We proceed to describe the structure of Iridium's messages, paying particular attention to the error correcting code which must be overcome to render the received packets undecodable.
In particular, we focus on Iridium Ring Alert (IRA) messages, which are the ones collected for the fingerprinting system being evaluated~\cite{smailes2023watch}.

\subsection{Structure of Iridium Ring Alert Messages}

In order to establish the power required to disrupt communication through jamming attacks, we must first understand the structure of the Iridium messages and their error correcting codes.
The Iridium Ring Alert (IRA) messages are designed to be received by all Iridium user terminals.
Each message contains information about the transmitting satellite, which includes its position, altitude, and a unique identifier for the satellite and its onboard transmitter.
These messages are preceded by a synchronization header, which is always identical for every transmission; the fingerprinter identifies the transmitter through small variations in this header.
The remainder of the message is $93$ bits in length, followed by a variable length set of data pages.
The $93$ bit section is protected by an interleaved ``BCH'' Error Correcting Code (ECC), which performs error detection and correction on the received bits -- we focus on this fixed section as we analyze the error correcting capacity of the code.
An overview of the precise operation of BCH codes was given by \textit{Walters et al.}~\cite{walters2020constant}.

Specifically, the ECC is applied to the message in three interleaved blocks of $21$ bits each.
Each $21$ bit block is protected by an additional $10$ parity bits to form an encoded $31$ bit block; the encoded message is given by multiplying the input message by the polynomial:\footnote{This ECC is implemented in the \textit{iridium-toolkit} software decoder, which can be found at the following URL: \url{https://github.com/muccc/iridium-toolkit}.}
$$
g(x) = x^{10} + x^7 + x^5 + x^4 + x^2 + x + 1
$$
The resulting code has minimum Hamming distance $5$; a hard decoder can therefore correct any $2$ bit error per block.
Therefore the probability of a block decode error for $p = \mathbb{P}(\mathit{bit~error})$, is given by:
$$
\mathbb{P}(\mathit{block~error}) = 1 - (1 - p)^{31} - 31p(1 - p)^{30} - {\binom{31}{2}}p^2\cdot(1 - p)^{29}
$$

The overall message is therefore decodable under this strategy if and only if there are no more than $2$ bit errors for every block within the message.
The probability of a message decode error, given input bit error $p$, is therefore given by:
$$
\mathbb{P}(\mathit{message~error}) = 1 - (1 - \mathbb{P}(\mathit{block~error}))^3
$$

Under this model, a 50\% message error rate is achieved at a bit error rate of $p \approx 0.08$.

\subsection{Related Work}\label{sec:related-work}

Some existing works look at attacks on fingerprinting systems, but the field has not yet been widely explored.
The primary focus of most of these works has been on impersonating device fingerprints: for example, the authors of~\cite{danevAttacks2010} use arbitrary waveform generators to replay messages with sufficient precision that the fingerprint is duplicated.
In~\cite{rehmanAnalysis2014} a similar approach is attempted using SDR hardware at lower sample rates, with limited success.
It has also been demonstrated that location fingerprinting techniques (using radio characteristics to determine the location of transmitters) can be disrupted through Gaussian noise~\cite{richterAttack2018}.
The authors of~\cite{tylerConsiderations2023} provide an overview of research that targets Specific Emitter Identification (SEI) systems (an alternative terminology for fingerprinting); all of these works focus on replay, impersonation, and circumvention of the system.
Finally, \cite{sunRobustness2022,liuRobust2023,karunaratnePenetrating2021} investigate attacks on SEI/fingerprinting systems in which an attacker causes misclassification of malicious messages through the use of adversarial machine learning.
To the best of our knowledge, denial-of-service attacks through jamming and disruption of transmitter fingerprinting systems have not yet been investigated.

\section{Threat Model}\label{sec:threat-model}

The objective of the attacker is to transmit radio interference, in the form of Additive White Gaussian Noise or a single tone, to cause errors in the fingerprinter which results in misclassification of the transmitter.
The desired outcome, from the attacker's perspective, is to cause misclassification even though the original signal remains decodable.
Therefore, as discussed in Section~\ref{sec:background}, we measure the robustness of the fingerprinting system by considering the attacker-to-victim power ratio required to achieve a given error rate on the decoder and fingerprinter systems.

We assume that the attacker has access to off-the-shelf software-defined radio hardware and a suitable amplifier to transmit in the correct frequency band.
The attacker's capabilities include maintaining presence within the vicinity of the receiver, such that the transmitted signal is received by the victim.
We further assume that the victim antenna is omnidirectional, the type of antenna used in real-world Iridium transceivers~\cite{beamcommunicationsIridium2017}.

\subsection{Required Hardware Budget}

We proceed to analyze the requirements of performing this attack with respect to two example transmitter systems; this allows us to estimate the hardware budget of the attacker, and understand the range over which the attacker can operate.

Firstly, the attacker uses a software-defined radio to emit either Gaussian noise or a single tone; no upconverter is required since Iridium operates at \SI{1600}{\mega\hertz}, which is well within the frequency range of most SDRs.
Secondly, the attacker requires a suitable amplifier, which can either be bought off-the-shelf, or made from components to reduce cost.
Finally, the attacker requires an antenna: we consider both an omnidirectional GPS patch antenna, which can deny service to receivers within a given area, and a high gain antenna which can target one particular receiver.
The full hardware components required to run the attack are tabulated in Table~\ref{tab:threat-model-components}.

It can be seen that, including the cost of the SDR, the equipment can cost as little as \qty{611}{\usd}; therefore deploying this attack is well within realm of a motivated hobbyist.

\subsection{Effective Attack Range}

\begin{table}
    \caption[Table of components required to execute a jamming attack on Iridium systems.]{Table of components required to execute a jamming attack on Iridium systems.\footnote{Prices are as recorded on 2024-01-12.}}
    \label{tab:threat-model-components}
    \centering
    \resizebox{\linewidth}{!}{
    \begin{tabular}{llll}
        \toprule
        \multicolumn{2}{l}{Component} & Cost (USD) & Power/Gain \\
        \midrule
        SDR & HackRF One & 340 & --- \\[.7em]
        Module amplifier & Mini Circuits ZHL-5W-2GX+ & 1195 & \qty{7}{\decibel\watt} \\
        RF IC amplifier & Qorvo QPA2237 & 155+100\footnote{The amplifier is sold as an integrated circuit, and will require installation into a custom-built circuit. We estimate this to cost approximately \qty{100}{\usd}.} & \qty{10}{\decibel\watt} \\[.7em]
        GPS patch antenna & Pulse 673-GPSGMSMA & 16 & \qty{0}{\decibel} \\
        High gain antenna & Chelton FPA21-16L/1258 & ---\footnote{The price for this component is not provided; it is designed for commercial use so it is likely to be expensive.} & \qty{21}{\decibel} \\
        \bottomrule
    \end{tabular}
    }
\end{table}

\begin{figure}
    \centering
    \includegraphics[width=\linewidth]{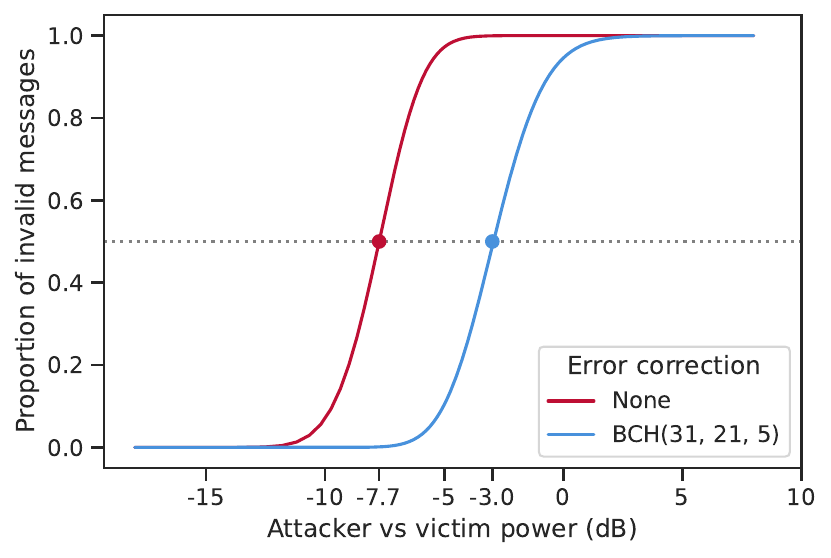}
    \caption{The proportion of Iridium messages which fail to decode as the jammer power increases, with and without the use of Iridium's built-in error correcting codes. A dashed line represents the point at which half of all messages fail to decode.}
    \label{fig:required-jammer-power}
\end{figure}

To better understand the threat caused by attackers of this class, we consider the distances over which jamming can be conducted on IRA messages with the hardware from the previous section.

Since IRA messages use Quadrature Phase Shift Keying (QPSK), it is established that the bit error probability corresponds to the \textit{energy per bit to noise power spectral density ratio}, $E_b/N_0$ by:
$$
\mathbb{P}(\mathit{bit~error})=0.5 \cdot \mathit{erfc}(\sqrt{E_b/N_0})
$$
where $\mathit{erfc}$ is the complementary error function derived from the cumulative Gaussian distribution.

Since QPSK encodes two bits per symbol, the jammer-to-signal power ratio is related to $E_b/N_0$ by:
$$
E_b/N_0 = \frac{1}{2}\cdot\frac{1}{P_a/P_v}
$$
Applying the ECC relationship between message error rate and bit error rate described in Section~\ref{sec:background}, in which $\mathbb{P}(\mathit{message~error}) = 0.5$ when $\mathbb{P}(\mathit{bit~error}) \approx 0.08$, the attacker-to-victim power ratio to cause a 50\% message error rate is given by:
$$
P_a/P_v = \frac{1}{2}\frac{1}{\mathit{erfc}^{-1}(2\cdot0.08)} = 0.503 = \qty{-2.98}{\decibel}
$$
This is illustrated in Figure~\ref{fig:required-jammer-power}, which also shows the lower power required to deny service if error correction is not used.

We can finally calculate the distances at which this $P_a/P_v$ can be achieved as distance varies, for our satellite system.
It has been previously determined in the context of radio overshadowing attacks that the peak received power of the Iridium-NEXT constellation is \SI{-145}{\deci\bel\watt}~\cite{salkield2023satellite}; therefore service is denied when the attacker's incident power is $-145 - 2.98 = -147.98$ \si{\deci\bel\watt}.
Assuming that the attacker has a clear line of sight to the victim, the attenuation of the attacker's signal over distance is given by the free space path loss formula:
$$
\mathit{fspl} = 20\log_{10}(d) + 20\log_{10}(f) - 27.55
$$
Where $d$ is distance in meters, and $f$ the frequency in \si{\mega\hertz}, $f \approx$ \SI{1600}{\mega\hertz} for Iridium.

\begin{figure}
    \centering
    \includegraphics[width=\linewidth]{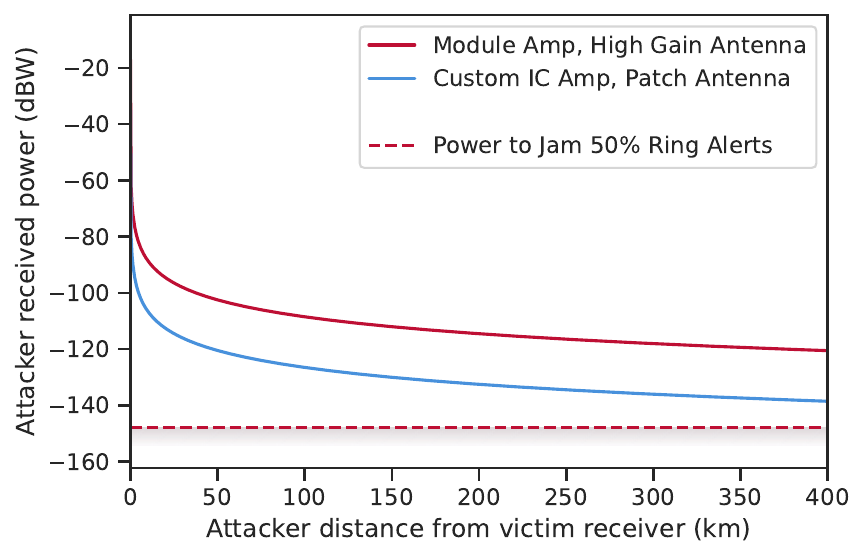}
    \caption{The received power of a noise jammer as distance to the victim antenna varies, under free space path loss. The dashed line represents the power required to cause a 50\% loss rate of Iridium Ring Alert messages; the region where any message loss is caused is shaded. It can be seen that with either set of equipment, line-of-sight attackers deny service over long distances.}
    \label{fig:equipment-path-loss}
\end{figure}

By applying this attenuation to the attacker's transmit power using the hardware given in Table~\ref{tab:threat-model-components}, we can determine the distances over which service can be denied.
The results are shown in Figure~\ref{fig:equipment-path-loss}, where it can be seen that either set of equipment is capable of denying services at very long distances.

It may seem surprising that communications can be jammed even at distances of hundreds of kilometers, but it should be considered that the attacker's \SI{10}{\deci\bel\watt} transmitter is competing with a \SI{9.5}{\deci\bel\watt} victim transmitter that itself undergoes \qty{781}{\kilo\meter} of path loss~\cite{salkield2023satellite}.
Although we still expect these values to be lower in a real-world setting where the attacker's signal is subject to multipath propagation and line-of-sight restrictions, this analysis nonetheless shows that cheaply available hardware is more than sufficient to deny service at long distances.

\section{Experiment Design}\label{sec:experiment-design}

\begin{figure*}
    \centering
    \includegraphics[width=.75\linewidth]{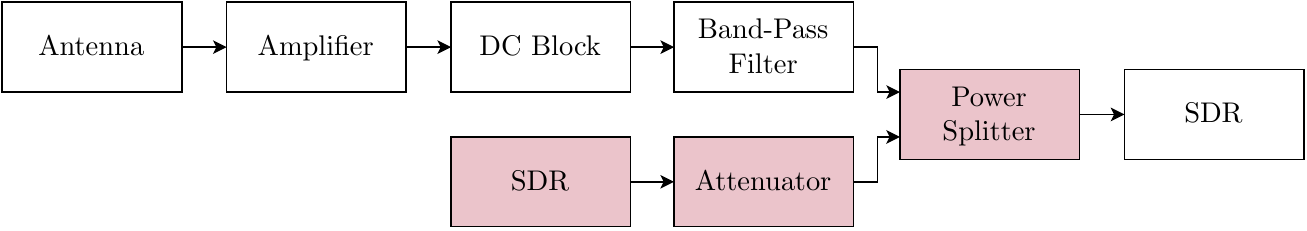}
    \caption{Overview of the hardware used to collect Iridium signals with additional noise. The hardware that has been added to enable variable noise injection has been highlighted.}
    \label{fig:data-collection-noise}
\end{figure*}

In this paper we test the hypothesis that satellite receiver systems can be more easily disrupted when fingerprinting systems are used to authenticate communication.
We achieve this by computing the required power for an attacker to disrupt the fingerprint of a legitimate signal through jamming, and comparing this to the power required to disrupt the contents of the message through conventional jamming techniques.
We have already established the attacker's constraints and ability to jam communication in Section~\ref{sec:threat-model}.
To assess the robustness of a fingerprinting system in the presence of the same interference, we next gather datasets of Iridium message headers with attacking interference added onto the signal.
We collected this data by adding noise onto real Iridium signals using radio hardware, and provide further datasets by adding synthetic interference to clean signals in software.

In this section we provide a summary of our data collection and processing pipeline, our software analysis, and an overview of the data collected.

\subsection{Data Collection}\label{sec:data-collection}

Our data collection setup is similar to the one used in our original work on \sysname{}~\cite{smailes2023watch}, with one notable difference: an additional transmitting SDR has been connected to the receiver, allowing interference to be added onto the incoming signal.
This setup is illustrated in Figure~\ref{fig:data-collection-noise}, and the hardware used is given in Table~\ref{tab:data-collection-hardware}.

\begin{figure}
    \centering
    \includegraphics[width=\linewidth]{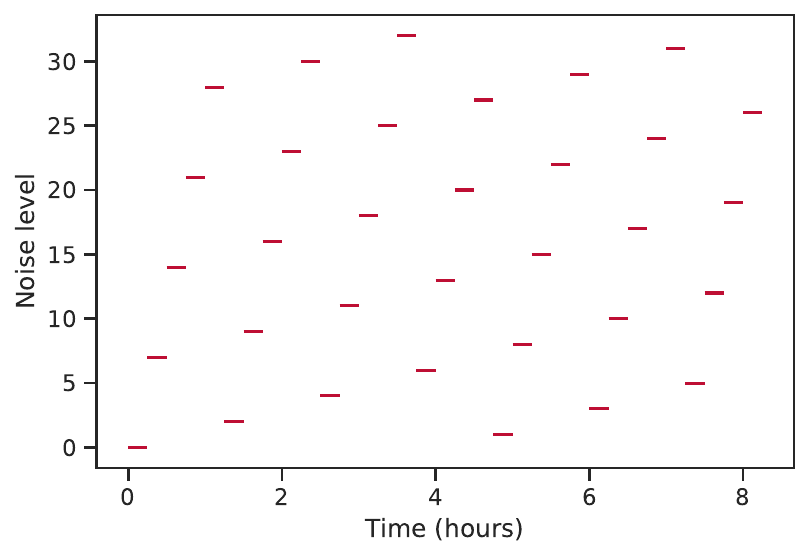}
    \caption{Level of noise added to the collected data over time. This pattern loops every \num{8}~hours.}
    \label{fig:interleave}
\end{figure}

In our previous work we note that the accuracy of the system is impacted by the time difference between the ``anchor'' message and the incoming message to be tested.
To remove this as a factor from our experiments, we interleave each of the noise levels -- this is illustrated in Figure~\ref{fig:interleave}.
Every \num{15}~minutes the data collection pipeline is restarted, and the noise level $N$ is incremented according to the formula $N = 7 * i \bmod 33$, where $i$ is the number of restarts.
This ensures an even spread of noise levels over time, and keeps the time gap between different portions of the dataset consistent.
A Gaussian noise function is used, generating signals with amplitude $N^2$.
Note that the amplitude of the noise here is arbitrary, and was selected to ensure a spread of values that covered the range our receiver was able to decode.
For our analysis in Section~\ref{sec:results} we look at the attacker-to-victim power ratio $P_a/P_v$, in decibels, computed from the received data.
This is calculated as follows:
$$
P_a/P_v = 20\log_{10}((\mathit{rms}_a - \mathit{rms}_v) / \mathit{rms}_v)
$$
Where $\mathit{rms}_a$ and $\mathit{rms}_v$ are the root-mean-squared amplitude of the attacker and victim signals respectively.\footnote{This assumes the attacker signal includes the victim signal. If working with a clean attacker signal, the ratio is instead $20\log_{10}(\mathit{rms}_a / \mathit{rms}_v)$.}

\begin{table}
    \caption[Hardware used for data collection.]{Hardware used for data collection.\footnote{Prices are as recorded on 2024-01-12.}}
    \label{tab:data-collection-hardware}
    \centering
    \begin{tabular}{llS[table-format=4.0]}
        \toprule
        Component           & Model number                    & {Cost (USD)} \\
        \midrule
        Antenna             & Iridium Beam RST740             & 1245 \\
        Low-noise amplifier & Mini-Circuits ZKL-33ULN-S+      &  209 \\
        DC block            & NooElec                         &   25 \\
        Band pass filter    & Mini-Circuits VBF-1560+         &   45 \\
        SDR (receiving)     & USRP N210, UBX 40 daughterboard & 5086 \\
        SDR (transmitting)  & BladeRF 2.0 micro xA4           &  540 \\
        Attenuator          & Mini-Circuits BW-S20W20+        &  174 \\
        Power splitter      & Mini-Circuits ZFRSC-123-S+      &   95 \\
        \bottomrule
    \end{tabular}
\end{table}

\subsection{Data Analysis}

\begin{figure}
    \centering
    \includegraphics[width=\linewidth]{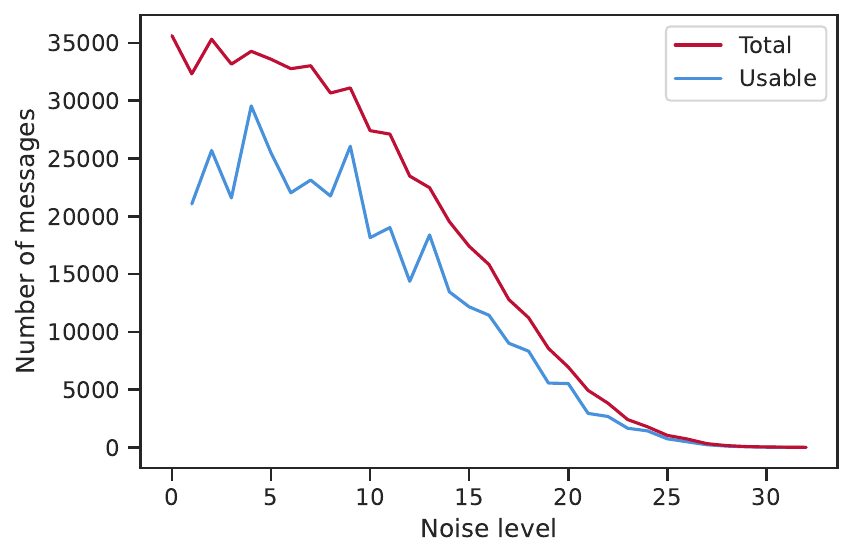}
    \caption{Number of messages collected for each noise level, and the number of messages that are useful for analysis (messages from a transmitter also seen in the zero-noise dataset).}
    \label{fig:num-messages-usable}
\end{figure}

\begin{figure*}
    \centering
    \includegraphics[width=\linewidth]{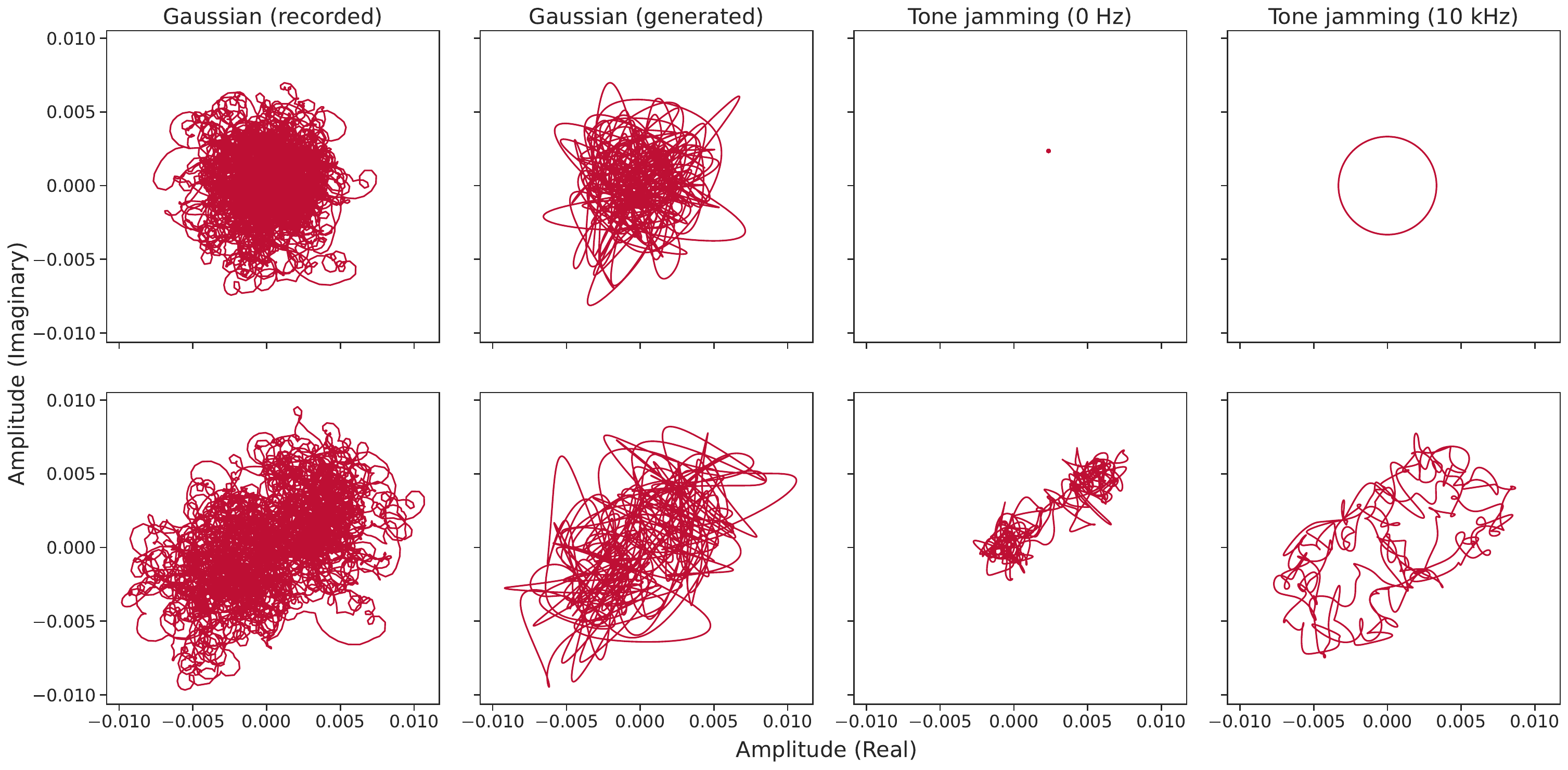}
    \caption{Comparison of the noise generated for each jamming technique, as constellation diagrams. The top row shows the jamming signal by itself, and the bottom row is the same jamming signal added onto a legitimate Iridium message header.}
    \label{fig:sample-noises}
\end{figure*}

\begin{figure}
    \centering
    \includegraphics[width=\linewidth]{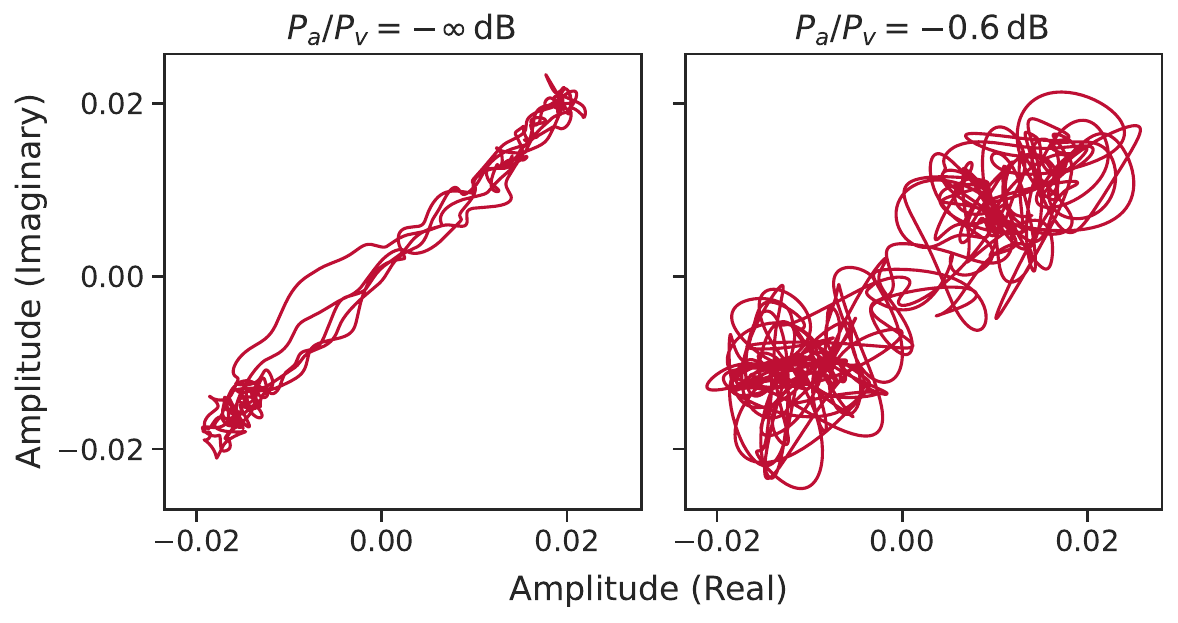}
    \caption{Two Iridium message headers received during data collection, depicted as constellation plots. The message on the left has no noise added, and the message on the right has Gaussian noise added, with an attacker-to-victim power ratio of \qty{-0.6}{\decibel}.}
    \label{fig:sample-multiple}
\end{figure}

We collected data using the above setup for \num{33}~days, during which \num{540066}~messages were received.
This dataset has been made openly available to aid future work.\footnote{The dataset can be found at \url{https://zenodo.org/records/10678124}, and the code at \url{https://github.com/ssloxford/SatIQ-noise}.}
For higher noise levels, fewer messages can be decoded; this is exacerbated by the software decoder's lack of error correction.
We can see in Figure~\ref{fig:num-messages-usable} that the number of received messages drops off smoothly as noise increases, and above a noise level of \num{32} (corresponding to $P_a/P_v = \qty{7.57}{\decibel}$) no further messages were received.
Also shown in this figure is the number of ``usable'' messages: the number of messages for which there exists at least one message from the same transmitter in the zero-noise control dataset.
Without such a message to compare against, fingerprinting cannot take place.
As data collection goes on, this will approach 100\% of the dataset.

In Figure~\ref{fig:sample-multiple} we see the effect of adding noise onto the message headers: it is still discernible as a PSK modulated signal, but the waveform is (unsurprisingly) significantly noisier.
Any of the original impairments on the signal which could have been used to identify the signal have likely disappeared into this noise.

\subsection{Software Jamming}\label{sec:experiment-software-jamming}

We also performed a software analysis, in which we added different noise and jamming signals to clean signals gathered from our data collection pipeline above.
This enabled us to evaluate a wider range of jamming techniques, and removed our reliance upon the decoder pipeline -- unlike the data collection above, the dataset does not shrink as we add more noise, since messages have already been demodulated and decoded.

We evaluate Gaussian noise jamming, as in the hardware experiments described above, and tone jamming, in which a constant frequency is added to the incoming signal.
These are the two main jamming techniques explored in recent works~\cite{martinelliRobustness2012}.
For our analysis we use generated Gaussian noise as well as Gaussian noise recorded from our data collection hardware, and tone jamming at different relative frequencies.

We can see each of these jamming techniques illustrated in Figure~\ref{fig:sample-noises}.
Each of these is calibrated to an attacker-to-victim ratio of \qty{-10}{\decibel}.
Note that the recorded and generated Gaussian noise look slightly different -- it is likely that the recorded Gaussian noise contains some higher frequencies outside the decoder's filter, resulting in a slightly different looking waveform.
We also illustrate two different forms of tone jamming: at a relative frequency of \qty{0}{\hertz}, the tone jamming appears as a constant offset, shifting the position of the victim signal by a fixed amount but not changing its shape.
At a higher relative frequency (\qty{10}{\kilo\hertz}) this instead looks like a circle, shifting different parts of the waveform by different amounts.

\section{Results}\label{sec:results}

In this section we assess the performance of the fingerprinting model against the noise and jamming signals collected in Section~\ref{sec:experiment-design}.
We use the code and model weights provided for \sysname{}~\cite{smailes2023watch}, using the \texttt{ae-triplet-final.h5} checkpoint.

For each noise level, we compare the fingerprints of received messages to the fingerprints of ``anchors'' in the zero-noise dataset with the same transmitter ID to provide a distance metric, telling us how similar the two message headers are.
If they are sufficiently different, the message will be rejected.
We use the distribution of distances at lower noise levels to establish a threshold at which we accept messages -- for this analysis we set this to be the 95th percentile.

We then find the attacker-to-victim power ratio at which 50\% of messages fall above this threshold, and take this to be a successful jamming of the fingerprinting system.
We compare this to the power required to disrupt communication using traditional methods, jamming the message through Gaussian noise.
This has already been established in Section~\ref{sec:threat-model} to be \qty{-2.98}{\decibel} for Iridium communication.

\subsection{Real-World Experiment}

\begin{figure}
    \centering
    \includegraphics[width=\linewidth]{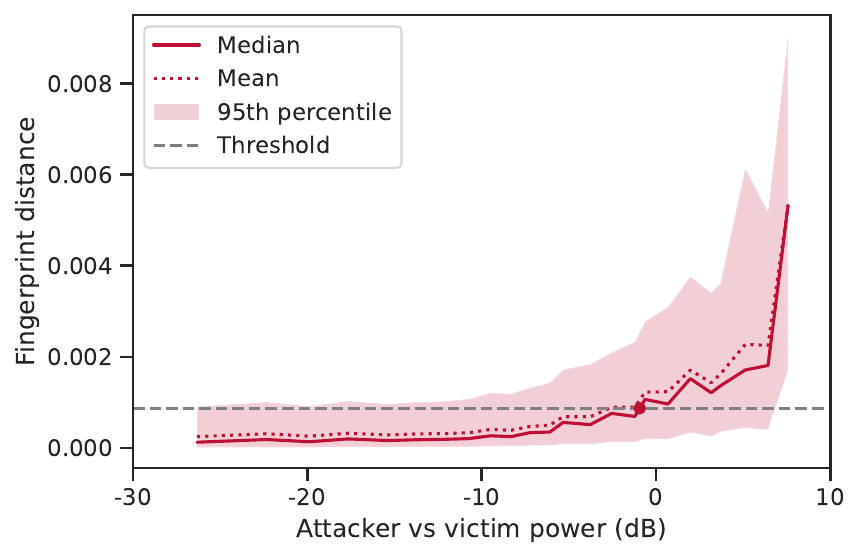}
    \caption{Distance reported by the \sysname{} fingerprinter for incoming messages under different levels of noise as captured by the real-world experimental setup, compared to anchor messages captured with no noise. A dashed line represents the fingerprinter threshold (set at the 95th percentile on the zero-noise dataset), below which messages are accepted by the system.}
    \label{fig:distances-hardware}
\end{figure}

We first look at the data collected using the hardware described in Section~\ref{sec:data-collection}, in which noise is added onto the incoming signals using hardware.
These results are summarized in Figure~\ref{fig:distances-hardware}.
Note the fluctuation at higher levels of attacker power -- this is caused by the smaller datasets captured for those levels, as the noise interferes with the decoding pipeline.
We can see that, as expected, adding noise to the signal disrupts the fingerprint.
The median fingerprint distances crosses the acceptance threshold at an attacker-to-signal ratio of \qty{-0.94}{\decibel}; at this point, half of all messages will be rejected by the fingerprinting system.
We also plot the mean for comparison purposes -- as noise increases, the variance of the distribution of fingerprints increases, bringing the mean distance up faster than the median distance.

This compares very favorably to the jamming analysis in Section~\ref{sec:threat-model}, in which we conclude that an attacker needs \qty{-2.98}{\decibel} to disrupt the message decoder through jamming.
Under this configuration, it takes more energy to disrupt the fingerprinter than it does to jam messages.

\subsection{Software Jamming}

\begin{figure}
    \centering
    \includegraphics[width=\linewidth]{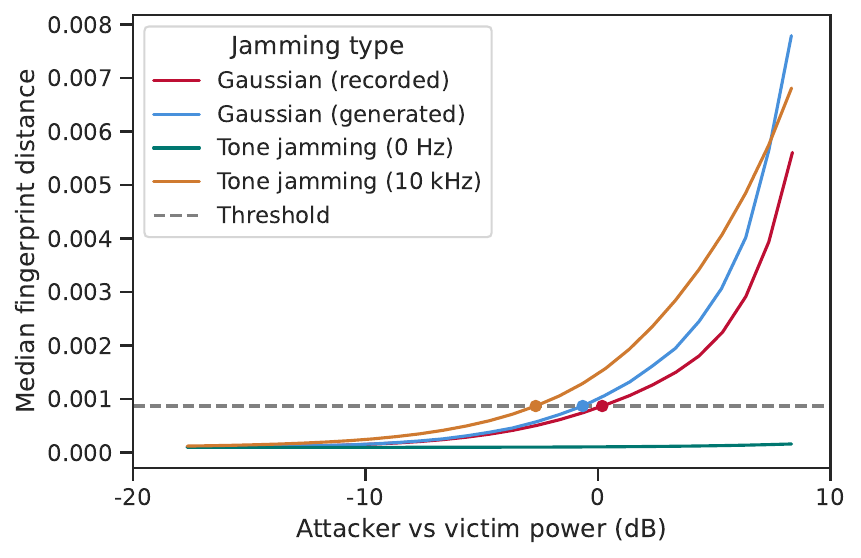}
    \caption{Distance reported by the \sysname{} fingerprinter for incoming messages with different levels of noise added in software, compared to anchors with no noise. The dashed line represents the fingerprinter threshold, accepting 95\% of messages when no noise is added.}
    \label{fig:distances-noisy}
\end{figure}

\begin{table}[t]
    \caption{Attacker-to-victim power ratio ($P_a/P_v$) required to disrupt 50\% of messages, under each jamming experiment.}
    \label{tab:thresholds}
    \centering
    \begin{tabular}{llS[table-format=2.4]}
        \toprule
        Target & Jamming type          & {$P_a/P_v$ (\unit{\decibel})} \\
        \midrule
        Decoder       & Gaussian              & -2.98 \\
        \arrayrulecolor{black!30}\midrule
        Fingerprinter & Gaussian (hardware)   & -0.94 \\
        \arrayrulecolor{black!30}\midrule
        Fingerprinter & Gaussian (recorded)   & 0.18 \\
        Fingerprinter & Gaussian (generated)  & -0.66 \\
        Fingerprinter & Tone jamming (0 Hz)   & 16.27 \\
        Fingerprinter & Tone jamming (10 kHz) & -2.68 \\
        \arrayrulecolor{black}\bottomrule
    \end{tabular}
\end{table}

\begin{figure}[t]
    \centering
    \includegraphics[width=\linewidth]{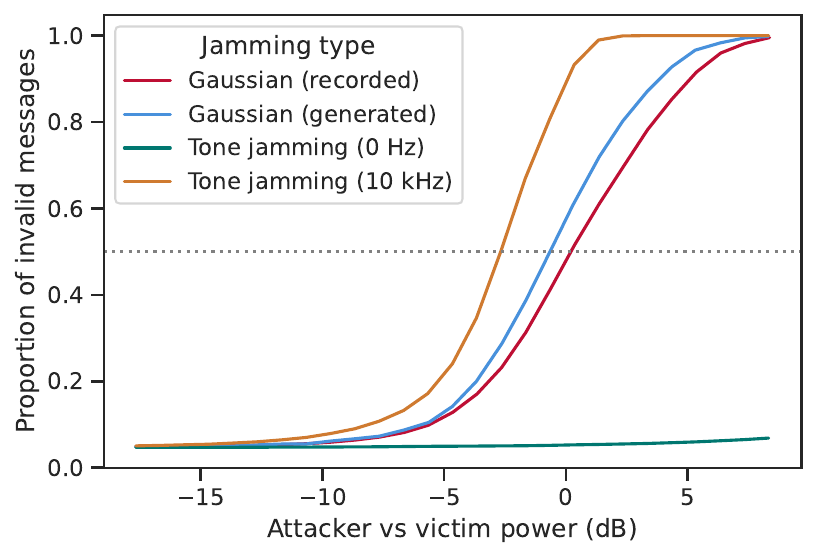}
    \caption{The proportion of victim messages rejected by the fingerprinting system as the attacker signal strength increases, under different jamming techniques.}
    \label{fig:distances-noisy-invalid}
\end{figure}

We also look at the jamming signals added in software in Section~\ref{sec:experiment-software-jamming}.
We can see the effect on the fingerprint as the attacker's transmit power increases in Figure~\ref{fig:distances-noisy}.
Additionally, the attacker-to-victim power required to disrupt 50\% of messages in this way can be seen in Table~\ref{tab:thresholds}.
The results for Gaussian jamming (both recorded and generated) are close to our experimental results, although slightly more attacker power is required than in the real-world evaluation.
This is likely due to the additional frequencies present in the Gaussian noise, outside the filter used by the decoder.

We can also see that tone jamming at a relative frequency of \qty{0}{\hertz} is ineffective; this is likely due to the heavy use of convolutional layers in the fingerprinting model, which are primarily concerned with relative changes in the waveform rather than absolute position.
However, we note that tone jamming at a relative frequency of \qty{10}{\kilo\hertz} is highly effective, successfully disrupting the fingerprinter at only \qty{-2.68}{\decibel} relative to the victim signal.

Finally, in Figure~\ref{fig:distances-noisy-invalid} we can see the proportion of messages rejected by the system as the attacker power increases, for each jamming technique.
We can see from this that tone jamming is significantly more effective at disrupting the fingerprint than Gaussian noise, disrupting over 90\% of messages at \qty{0}{\decibel} relative to the victim signal, and reaching 100\% at approximately \qty{2.5}{\decibel}.

\section{Discussion}\label{sec:discussion}

In Section~\ref{sec:threat-model} we look at the error rate of the Iridium decoder under Gaussian jamming, as the received attacker power increases.
In Figure~\ref{fig:distances-noisy-invalid-with-decoder} we compare this to the fingerprinter error under each of the most effective jamming techniques, from both our hardware experiment and software analyses.\footnote{Note that the baseline error rate for the fingerprinter depends upon the threshold at which messages are accepted. In this case we set the threshold at the 95th percentile, so the base error rate is $0.05$.}
We can see from this figure that under every form of jamming tested, it takes a greater amount of attacker power to disrupt the fingerprint than to disrupt the message through jamming.
In all cases, the difference in required attacker power is quite small, within \qty{2.5}{\decibel}.
Tone jamming also requires much less power to disrupt the fingerprint than Gaussian jamming, getting much closer to the power level required for jamming.
under Gaussian jamming, the decoder error rate is greater than the fingerprinting system's message rejection rate as the attack power increases.
We have therefore refuted our initial hypothesis, showing that the use of fingerprinting to authenticate satellite communication does not significantly weaken the system against jamming attacks.

There are a range of potential explanations for why this is the case.
First and foremost, fingerprinting in the context of satellite systems is much more difficult than terrestrial systems, due to the high levels of free space path loss and atmospheric distortion; any identifiable information present in the signal will be subject to significant atmospheric attenuation.
In order to achieve good performance, the fingerprinting model must be able to overcome this, and it is likely that in the process of doing so it becomes good at ignoring Gaussian noise -- this would be unsurprising, as convolutional autoencoders (the architecture used by this model) have been demonstrated in other works to be effective at removing noise in a range of applications~\cite{ahmedMedical2021,bhowickStacked2019}.
The fingerprint therefore only fails to be identified when the injected noise is so high that the message itself can no longer be decoded.
It may be the case that fingerprinting systems trained on terrestrial data are less robust against noise -- they do not need to perform nearly as much noise reduction as satellite fingerprinters, so a smaller amount of noise could be required to disrupt them.

\begin{figure}[t]
    \centering
    \includegraphics[width=\linewidth]{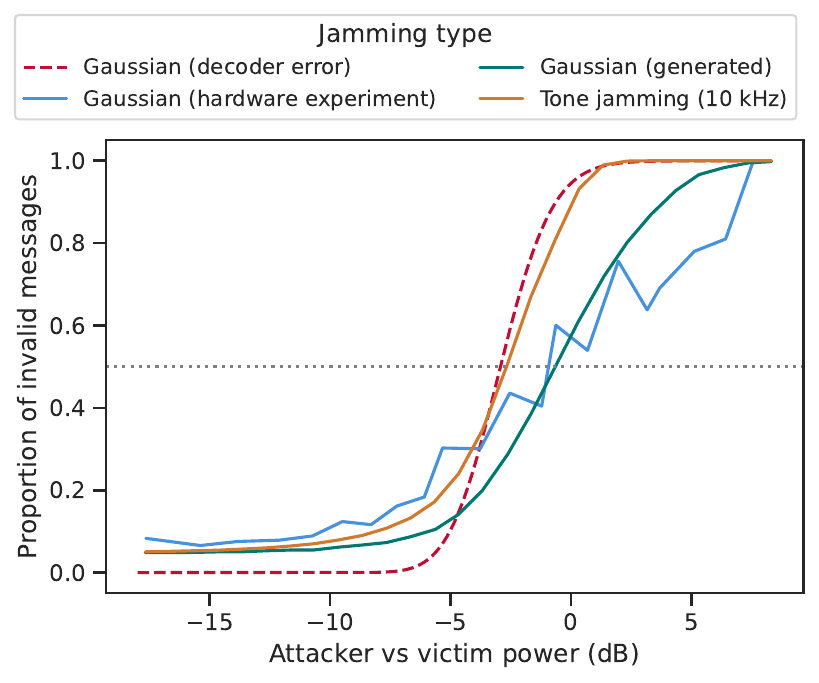}
    \caption{The proportion of invalid messages as received attacker power increases, comparing the theoretical decoder error (dashed line) to the fingerprinter error under the best-performing experimental configurations.}
    \label{fig:distances-noisy-invalid-with-decoder}
\end{figure}

It is also important to note that the decoder and fingerprinter are fundamentally operating on different parts of the signal.
The fingerprinter operates on the message header only -- this portion of the message is unchanging, and designed to be robust and easy to correlate over.
On the other hand, the decoder primarily operates on the message body, which changes between messages and is generally expected to fail first in the presence of noise -- enough noise needs to be added to introduce only a small number of bit errors, which may not destroy all identifying information present in the message header.

It is possible that an attacker aiming to specifically disrupt the fingerprinting system may be able to craft a waveform that disrupts the fingerprinter even more effectively, with lower transmit power.
There is a large body of existing research that has demonstrated the effectiveness of adversarial machine learning techniques to execute these types of attacks, in systems ranging from image classifiers~\cite{machadoAdversarial2021} to network intrusion detection systems~\cite{alatwiAdversarial2021}.
There has also been some work investigating terrestrial RF fingerprinting systems, with theoretical analyses~\cite{liuRobust2023} and small-scale experimentation~\cite{karunaratnePenetrating2021}.
Future work may consider evaluating this type of attack on satellite fingerprinting systems.
Such techniques usually require black-box access to the target system, if not access to the underlying model weights, and are highly targeted to a single system.
We therefore do not consider them in this work, instead considering an attacker employing the use of standard jamming techniques, and concluding that the use of fingerprinting does not open up significant new weaknesses in this case.

\section{Conclusion}\label{sec:evaluation}

In this paper we have provided a novel analysis of the vulnerability of satellite fingerprinting systems against Gaussian and tone jamming interference.
We show that the~\sysname{} fingerprinting system is not significantly more vulnerable to this interference than the Iridium message decoder itself, so fingerprinting can be used without opening the system up to easy denial-of-service attacks.

Existing legacy satellite systems are likely to remain in use for years or decades to come, and in the absence of cryptographic authentication, fingerprinting techniques can provide confidence that downlinked data is legitimate and has not been spoofed or tampered with.
This provides an increased degree of trust in the systems, enabling them to remain useful for the remainder of their operational period.

\section*{Acknowledgments}
We would like to thank armasuisse Science + Technology for their support during this work.
Joshua was supported by the Engineering and Physical Sciences Research Council (EPSRC).
Sebastian was supported by the Royal Academy of Engineering and the Office of the Chief Science Adviser for National Security under the UK Intelligence Community Postdoctoral Research Fellowships programme.

\printbibliography

\end{document}